\documentclass[11pt]{article} 

\usepackage{mathtools, amsfonts, amsthm, latexsym, amssymb,dsfont}
\usepackage[T1]{fontenc}
\usepackage[tracking=true]{microtype}
\usepackage{lmodern}

\usepackage[margin=1in]{geometry}

\usepackage{stmaryrd}
\usepackage{amsmath}
\usepackage{amsfonts}
\usepackage{amssymb}

\usepackage{enumitem}
\usepackage{graphicx}
\usepackage{subcaption}
\usepackage{cite}

\usepackage{color}
\definecolor{darkblue}{cmyk}{0.9,0.9,0,0}
\definecolor{wine-stain}{rgb}{0.5,0,0}
\usepackage[colorlinks, allcolors=darkblue, linktoc=all]{hyperref} 
\usepackage{here}

\def \be {\begin{equation}}
\def \ee {\end{equation}}

\renewcommand{\O}{\mathcal{O}}

\newcommand{\ket}[1]{\vert #1 \rangle}

\newcommand{\matrixel}[3]{\langle #1 \vert #2 \vert #3 \rangle}
\newcommand{\vev}[1]{{\langle #1 \rangle}}

\begin{document}
\thispagestyle{empty}
\vspace*{1in} 

\begin{center}
{\Large \textbf{The large charge expansion and AdS/CFT}}
\vspace{.6in}

\textbf{Anton de la Fuente\textsuperscript{a,b} and Jann Zosso\textsuperscript{c}}
\vspace{.2in} 

{\small \textit{\textsuperscript{a}Kavli Institute for the Physics and Mathematics of the Universe, The University of Tokyo, Kashiwa, Japan}} \\
{\small \textit{\textsuperscript{b}Theoretical Particle Physics Laboratory (LPTP), Institute of Physics, EPFL,  Lausanne, Switzerland}} \\
{\small \textit{\textsuperscript{c}Institute for Theoretical Physics, ETH Zurich, Wolfgang-Pauli-Strasse 27, 8093, Zurich, Switzerland}}
\end{center}
\vspace{.2in}

\begin{abstract}
The scaling dimensions of charged operators in conformal field theory were recently computed in a large charge expansion. We verify this expansion in a dual AdS model.
Specifically, we numerically construct solitonic boson star solutions of Einstein-Maxwell-Scalar theory in AdS$_4$ and find that its mass at large charge reproduces the universal form of the lowest operator dimension in the large $U(1)$ charge sector of the dual 2+1 dimensional CFT.
\end{abstract}

\setcounter{page}{0}

\newpage

%%%%%%%%%%%%%%%%%%%%%%%%%%%%%%%%%%%%%%%%%%%%%%%%%%%%%%
%%%%%%%%%%%%%%%%%%%%%%%%%%%%%%%%%%%%%%%%%%%%%%%%%%%%%%

\section{Introduction}

The most successful methods in quantum field theory involve a perturbation expansion in a small parameter. Recently \cite{origlQ, alvarezGaume, lQ}, a new such parameter was found in conformal field theory~(CFT). When two of the operators in a correlation function have a large global charge, the correlator may be computable in a large charge expansion. 

Consider a 2+1 dimensional CFT with a $U(1)$ global symmetry. Let $\Delta$ be the scaling dimension of the lowest dimension operator of charge $Q$. The large charge expansion predicts
\begin{equation} \label{prediction}
	\Delta = Q^{3/2}  \sum_{n=0}^\infty \frac{c_n}{Q^n} + \text{calculable},
\end{equation}
where the $c_n$ are theory dependent Wilson coefficients and ``calculable'' can be viewed as universal quantum corrections (giving rise to precise numerical predictions \cite{lQo4, epsilon}). In this note, we verify the form of \eqref{prediction} in a simple model using AdS/CFT.

The large $Q$ expansion makes use of the state-operator correspondence. Let $\ket Q$ be the state corresponding to the operator under consideration. By construction, $\ket Q$ has a non-vanishing $U(1)$ charge density. Furthermore, it is assumed that in the infinite volume limit at fixed charge density, the $U(1)$ symmetry is spontaneously broken. Such a state is known as a superfluid \cite{son}. We will construct a simple model in AdS$_4$ that is dual to a superfluid in CFT$_3$.
% at zero entropy?

By the AdS/CFT correspondence:
\begin{enumerate} 
\item The stress tensor of the CFT is dual to a dynamical metric with asymptotically AdS boundary conditions.
\item The global $U(1)$ current $J$ of the CFT is dual to a $U(1)$ gauge field $A$ in AdS.
\item The non-vanishing charge density imposes the boundary condition $A^0 \to \mu + z \matrixel{Q}{J^0}{Q}$, where~$z$ is the radial coordinate and $\mu$ is the chemical potential.
\end{enumerate}
Physically, the boundary condition on $A$ imposes a non-vanishing radial electric field at the boundary. Any bulk solution with such an electric field will be dual to a CFT state of large charge. However, in order to describe a superfluid, we will restrict ourselves to states that spontaneously break the global $U(1)$ symmetry. This means that in the CFT, there is an operator $\O$ with $\vev \O \neq 0$. In AdS, 
\begin{enumerate}[resume]
\item $\O$ is dual to a field $\phi$ with the boundary condition $\phi \to z^\Delta \vev \O + z^{d-\Delta} \times \text{source}$. Since the symmetry is spontaneously broken,  the source term vanishes.
\end{enumerate}
A minimal model with these properties is Einstein-Maxwell-Scalar theory on asymptotically global AdS. We will numerically construct solitonic boson star solutions of this model and find that its mass at large charge reproduces \eqref{prediction}.

Many aspects of Einstein-Maxwell-Scalar theory  were previously studied in \cite{A50,A5I,A5II,A4I,A4II,A40}. Numerics were also employed using the Poincar\'{e} patch counterpart to this model in the study of holographic superconductors \cite{0,XI,VIII,VI}.

%%%%%%%%%%%%%%%%%%%%%%%%%%%%%%%%%%%%%%%%%%%%%%%%%%%%%%
\section{Setup}
We consider
\be
\label{action}
S = \tfrac{1}{16\pi G_4}\int d^4x \sqrt{-G}\left[\tfrac{6}{L^2}  + \mathcal{R}-\tfrac{1}{4}F_{a b}F^{a b}-(D_a\phi)^*D^a\phi + \tfrac{2}{L^2}|\phi|^2 \right],
\ee
where $G_4$ is the gravitational constant, $\mathcal{R}$ is the Ricci scalar, and $L$ denotes the AdS radius.\footnote{The equations of motion are invarant under the scaling 
$
r\shortrightarrow \lambda r \, ,\;  t \shortrightarrow \lambda t \, ,\; L\shortrightarrow \lambda L \, ,\;  q\shortrightarrow q/ \lambda
$
which we shall used to set $L=1$.}
The covariant derivative is defined as $D_a\phi=\partial_a\phi-iqA_a\phi$, where $q$ is the charge of the scalar field.

We are looking for static and spherically symmetric solutions and therefore choose the following ansatz for the bulk metric,
\begin{gather}
\label{metric}
ds^2=-g(r) b(r)^2 dt^2 +\frac{dr^2}{g(r)}+r^2 d\theta^2+r^2 \sin^2\theta\, d\varphi^2,
\end{gather}
where $b \to 1$ and  $g\to \frac{r^2}{L^2}+1$ as $r\to \infty$ for asymptotically AdS boundary conditions.

Spherical symmetry, time-independence, radial gauge, and the restriction to electrically charged solutions allows the following ansatz for the fields,
\be
\label{fansatz}
\phi=\phi(r) \qquad A=a(r) dt.
\ee
Moreover, the $r$-component of Maxwell equations implies that the phase of $\phi$ must be constant, allowing us to take $\phi$ to be real without loss of generality.

The equations of motion result in four coupled differential equations and the following asymptotic behaviour of the fields,
\begin{align}
\phi(r\shortrightarrow\infty)&=\frac{\phi_1}{r}+\frac{\phi_2}{r^2}+O\left(1/r^3\right) \\
a(r\shortrightarrow\infty)&=\mu-\frac{\rho}{r}+O(1/r^2)\\
g(r\shortrightarrow \infty)&=r^2+1+\frac{\phi_1^2}{2}-\frac{m}{r}+O\left(1/r^2\right) .\label{labsefalloff}
\end{align}
To impose that the dual CFT state is a superfluid with zero entropy, the solutions we consider have no horizon, vanishing~$\phi_1$, and non-vanishing $\mu$, $\rho$, and $\phi_2$.

The parameter $\rho$ is then related to the total charge $Q_{\text{AdS}}$ of the solution through Gauss' law, which in our convention is 
\be 
Q_{\text{AdS}}=\frac{L^2}{16\pi G_4}\,\lim_{r\rightarrow \infty}\,\int_{S^2}\star F=\frac{L^4}{4G_4}\,\rho\, ,
\ee
where $S^2$ is a two-sphere at constant $t$ and $r$ and $\star F$ is the Hodge star dual of the field strength. Similarly, the AdS invariant boundary conditions allow us to make the link between the physical mass M of the solution and the coefficient $m$ in the fall-off of the lapse function \eqref{labsefalloff}
\be
M=\frac{L^4}{2G_4}m\,.
\ee

The AdS/CFT dictionary relates the boundary values $m$ and $\rho$ to the dimension $\Delta_Q/l$ and charge $Q$ of the CFT respectively. Following \cite{A40} we will choose units in which $2 G_4=1$, such that\footnote{A change to Fefferman-Graham coordinates shows that one can choose the AdS radius to coincide with the radius of the cylinder $L=l=1$.}
\be 
\label{massrel}
M=m=\Delta_Q\quad\text{and}\quad Q_{\text{AdS}}=\tfrac{1}{2}\,\rho=Q\,.
\ee

%%%%%%%%%%%%%%%%%%%%%%%%%%%%%%%%%%%%%%%%%%%%%%%%%%%%%%

\section{Results}
\label{soli}

\begin{figure}[H]
  \centering
 \includegraphics[width=.45\linewidth]{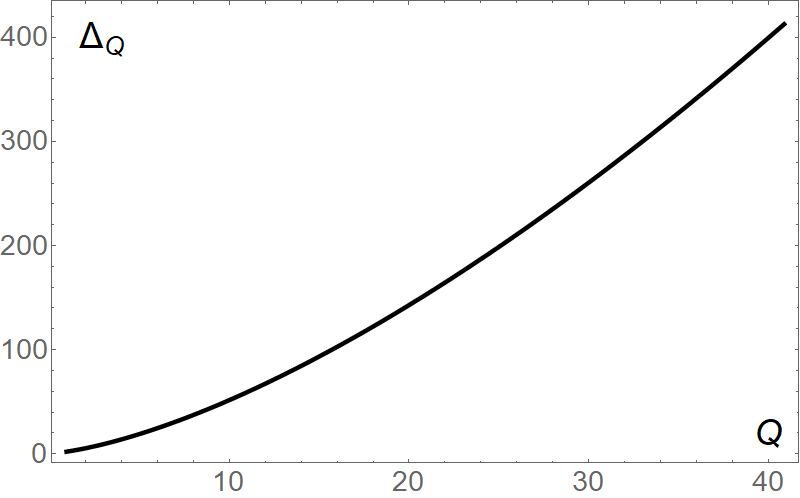}
\caption{ The numerical result of $M(Q_{\text{AdS}})=\Delta_Q(Q)$ along the soliton branch $q^2=1.4$.}
\label{q_1} 
\end{figure}
We numerically integrated the EOMs using a shooting method and constructed horizon-less solutions with non-vanishing scalar, electric, and gravitational fields. Such solutions are often referred as ``boson stars''  \cite{A50,A4I,A5I,A4II,A5II} and consists of a one parameter family for each value of the scalar charge~$q$. 
For concreteness, we take $q^2=1.4$ and plot $\Delta_Q$ in Figure~\ref{q_1}. Assuming the expansion~\eqref{prediction}, a fit of the numerical data gives the following,
\be 
\label{fit}
\Delta_Q^{\text{fit}}=1.56369\;Q^{3/2}+0.68411\;Q^{1/2}-0.03174\; \frac{1}{Q^{1/2}}.
\ee

\begin{figure}[H]
\centering
\hspace{0.01\linewidth}
\begin{subfigure}{0.46\linewidth}
  \centering
  \includegraphics[width=\linewidth]{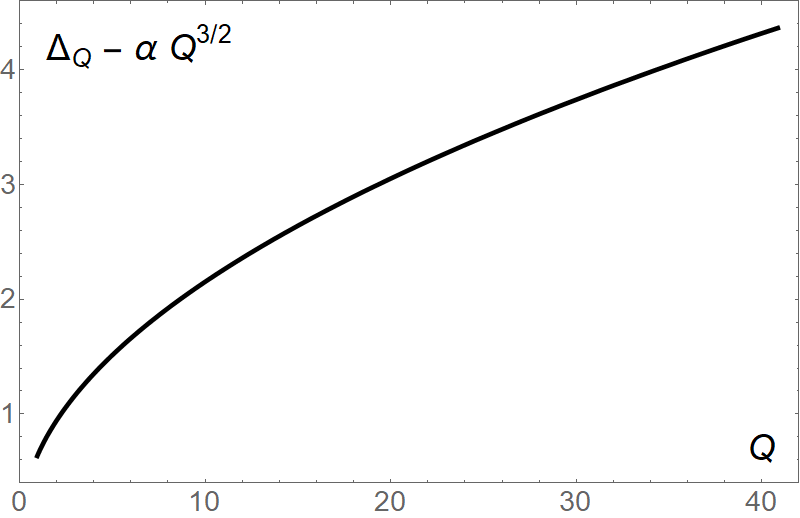}
\caption{}
\label{q_2}
\end{subfigure}
\hspace{0.02\linewidth}
\begin{subfigure}{0.48\linewidth}
  \centering
\includegraphics[width=\linewidth]{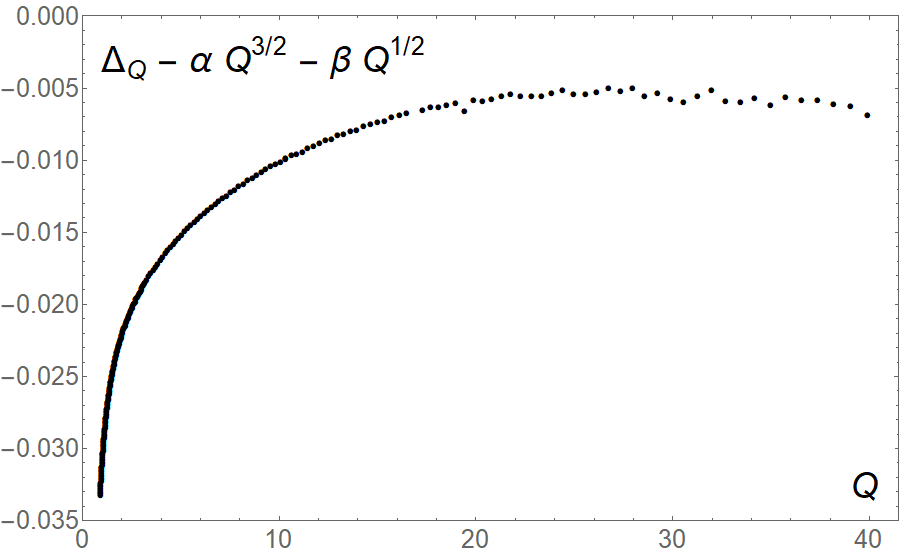}
\caption{}
\label{q_3}
\end{subfigure}
\caption{(a) Subtracting the first term of the fit \eqref{fit} from the data reveals a $\sim Q^{1/2}$ behavior, confirming the subleading term in \eqref{prediction}. (b) Subtracting also the $\beta$-term of \eqref{fit} reveals the $\sim\, Q^{-1/2}$ behavior.}
\label{q}
\end{figure}

Next, we plot the difference between the numerical values of $\Delta_Q$ (Figure \ref{q_1}) and the terms of $\Delta_Q^{\text{fit}}$ \eqref{fit} in Figures~\ref{q_2} and \ref{q_3}. Fitting the curves \ref{q_2} and \ref{q_3} with a functions of the form $b\,Q^{a}$ yields $a_b\simeq 0.510$ and $a_c\simeq -0.506$ respectively, clearly showing the subleading and the second order behaviours $\sim Q^{1/2}$ and $\sim Q^{-1/2}$.
Figure \ref{q_3} indicates that the precision of our code is starting to reach its limits at the second order. However, the agreement with \eqref{prediction} is still clearly visible. We therefore confirm the predicted universal form of the scaling dimension $\Delta_Q$ \eqref{prediction} of the operator of lowest dimension with charge $Q$.

%%%%%%%%%%%%%%%%%%%%%%%%%%%%%%%%%%%%%%%%%%%%%%%%%%%%%%
%%%%%%%%%%%%%%%%%%%%%%%%%%%%%%%%%%%%%%%%%%%%%%%%%%%%%%

\section{Discussion}
\label{discussion}
We numerically constructed boson star solutions in a simple AdS model and computed their mass as a function of their charge. We showed that the result is consistent with the prediction \eqref{prediction}.
Our computation was purely classical and therefore did not reproduce the ``calculable'' terms in~\eqref{prediction}. To reproduce those, we would need to include quantum fluctuations about the boson star background. Such fluctuations would be dual to the fluctuations of the goldstone boson in the CFT \cite{hartnoll}. We note that higher order terms in the action~\eqref{action} will presumably only change the numerical values of the coefficients $c_n$ in the expansion~\eqref{prediction}. 

Apart from boson stars, the theory \eqref{action} also admits other charged solutions, such as black holes with scalar hair. See \cite{phase} for a phase diagram. However, such black holes do not correspond to superfluids because they have a large zero-temperature entropy. Furthermore, the Reisner-Nordstrom black hole considered in \cite{reffert} does not break the $U(1)$ symmetry and therefore does not have the correct symmetry breaking pattern of a superfluid. 

In future work, we would like to compute the spectrum of small fluctuations as well as boundary correlation functions. This would allow us to further check the predictions of \cite{origlQ, lQ} and to make contact with the bootstrap analysis of \cite{resumlQ}.

%%%%%%%%%%%%%%%%%%%%%%%%%%%%%%%%%%%%%%%%%%%%%%%%%%%%%%
%%%%%%%%%%%%%%%%%%%%%%%%%%%%%%%%%%%%%%%%%%%%%%%%%%%%%%

\section*{Acknowledgments}
We thank Gabrielle Cuomo, Simeon Hellerman, Orestis Loukas, Alexander Monin, Domenico Orlando, Joao Penedones, Riccardo Rattazzi, Suzanne Reffert for discussions and conversations. The work of A.D.\ is partially supported by the Swiss National Science Foundation under contract 200020-169696, the National Center of Competence in Research SwissMAP, and the World Premier International Research Center Initiative, MEXT, Japan.

%%%%%%%%%%%%%%%%%%%%%%%%%%%%%%%%%%%%%%%%%%%%%%%%%%%%%%
%%%%%%%%%%%%%%%%%%%%%%%%%%%%%%%%%%%%%%%%%%%%%%%%%%%%%%

\bibliographystyle{utphys}
\bibliography{references} 
\end{document}